\documentclass[article,aps,twocolumn,showpacs,amsmath,amssymb,superscriptaddress,nofootinbib,floatfix]{revtex4-2}

\usepackage{graphicx}
\usepackage{dcolumn}
\usepackage{bm}
\usepackage[colorlinks=true,linkcolor=blue,citecolor=blue,urlcolor=blue]{hyperref}

\usepackage{array,booktabs,tabularx}
\usepackage[caption=false]{subfig}
\usepackage[USenglish]{babel}
\usepackage{braket}
\usepackage{xcolor}
\usepackage{amsfonts}
\usepackage{amssymb}
\usepackage[normalem]{ulem}
\usepackage{amsmath}
\usepackage[utf8]{inputenc}

\begin{document}


\title{Flat band driven competing charge and spin instabilities in the altermagnet CrSb}

\author{A. Korshunov}
\thanks{These authors contributed equally.}
\affiliation{Donostia International Physics Center (DIPC), San Sebastián, Spain}

\author{M. Alkorta}
\thanks{These authors contributed equally.}
\affiliation{Centro de Física de Materiales (CFM-MPC), CSIC-UPV/EHU, San Sebastián, Spain}
\affiliation{Fisika Aplikatua Saila, Gipuzkoako Ingeniaritza Eskola, University of the Basque Country (UPV/EHU), San Sebastián, Spain}

\author{C.-Y. Lim}
\affiliation{Donostia International Physics Center (DIPC), San Sebastián, Spain}

\author{F. Ballester}
\affiliation{Donostia International Physics Center (DIPC), San Sebastián, Spain}
\affiliation{Fisika Aplikatua Saila, Gipuzkoako Ingeniaritza Eskola, University of the Basque Country (UPV/EHU), San Sebastián, Spain}

\author{Cong Li}
\affiliation{Department of Applied Physics, KTH Royal Institute of Technology, Stockholm 11419, Sweden}
\affiliation{Beijing National Laboratory for Condensed Matter Physics, Institute of Physics, Chinese Academy of Sciences, Beijing 100190, China} 

\author{Zhilin Li}
\affiliation{Beijing National Laboratory for Condensed Matter Physics, Institute of Physics, Chinese Academy of Sciences, Beijing 100190, China}

\author{D. Chernyshov}
\affiliation{Swiss-Norwegian BeamLines at European Synchrotron Radiation Facility, BP 220, F-38043 Grenoble Cedex, France}

\author{A. Bosak}
\affiliation{European Synchrotron Radiation Facility (ESRF), BP 220, F-38043 Grenoble Cedex, France}

\author{M. G. Vergniory}
\affiliation{Donostia International Physics Center (DIPC), San Sebastián, Spain}
\affiliation{Département de Physique et Institut Quantique, Université de Sherbrooke, Sherbrooke, Québec, Canada}

\author{Ion Errea}
\email{ion.errea@ehu.eus}
\affiliation{Donostia International Physics Center (DIPC), San Sebastián, Spain}
\affiliation{Centro de Física de Materiales (CFM-MPC), CSIC-UPV/EHU, San Sebastián, Spain}
\affiliation{Fisika Aplikatua Saila, Gipuzkoako Ingeniaritza Eskola, University of the Basque Country (UPV/EHU), San Sebastián, Spain}

\author{S. Blanco-Canosa}
\email{sblanco@dipc.org}
\affiliation{Donostia International Physics Center (DIPC), San Sebastián, Spain}
\affiliation{IKERBASQUE, Basque Foundation for Science, 48013 Bilbao, Spain}

\date{\today}

\begin{abstract}
The confinement of electronic wavefunctions in momentum space can give rise to flat electronic bands, where the quenching of kinetic energy enhances the density of states and amplifies interaction effects. Such conditions are fertile ground for emergent quantum phases, as spin, charge and lattice degrees of freedom become strongly entangled. In these regimes, subtle competitions between intertwined order parameters often dictate the macroscopic ground state, producing complex and sometimes unexpected collective behavior. Here we show that the altermagnet CrSb provides a realization of this scenario, and uncover short-range charge-order fluctuations at the M point of the Brillouin zone, q$^*$=($\frac{1}{2}$\ 0), persisting above the Néel temperature (T$_\mathrm{N}$). Remarkably, these fluctuations collapse upon entering the magnetically ordered phase, revealing a direct and robust competition between charge and spin order. At T$_\mathrm{N}$, the phonon dispersion at q* develops a pronounced Kohn-like anomaly, signaling strong electron–phonon coupling in the vicinity of the magnetic transition. Below T$_\mathrm{N}$, exchange striction dramatically renormalizes the associated soft phonon mode by approximately $\sim$6 meV, the largest spin-phonon coupling ever reported. First-principles calculations attribute this behavior to a  strong coupling between nearly dispersionless electronic states and a phonon branch that appears unstable at the harmonic level only when no magnetic order is considered, revealing the large sensitivity of the lattice to magnetic symmetry breaking. The competition between charge and spin order parameters, amplified by flat-band physics, drives the observed phonon anomaly and its abrupt reconstruction at T$_\mathrm{N}$. With its chemically simple structure and symmetry-protected altermagnetic state, CrSb emerges as a model platform to explore how flat electronic bands mediate giant spin–phonon coupling and competing broken symmetries. 
\end{abstract}

\maketitle

Correlated quantum materials are distinguished by a landscape of nearly degenerate ground states in which multiple symmetry-breaking instabilities compete and coexist, reorganizing the entire electronic structure and phonon spectrum \cite{Balents_2010,Keimer_2015}. The resulting broken-symmetry phases encode a complex entanglement of spin, charge, orbital and lattice degrees of freedom \cite{Khomskii_1997}, often governing the macroscopic ground state and low energy collective excitations. Canonical examples include charge-density waves (CDWs) and spin-density waves (SDWs), which break translational and time-reversal symmetry \cite{Gruner_1994}, respectively, and reconstruct the Fermi surface (FS). Yet some of the most orders remain partially hidden -- elusive in conventional probes but profoundly shaping the phase diagram. Such intertwined or “vestigial” states have emerged as central themes in heavy-fermion compounds \cite{Riggs_2015,Chandra_2002,Villaume_2008}, kagome metals \cite{Guguchia_2023,Mielke_2022,Saykin_2023} and high-temperature superconductors \cite{Chakra_2001,Wu_2011,Fernandes_2014,Frachet_2020,Ghiringhelli_2012,Chang_2012}, where bandwidth-controlled instabilities and competing symmetry breakings generate rich phase diagrams and unconventional collective phenomena.

\begin{figure*}
    \centering
    \includegraphics[width=1.0\linewidth]{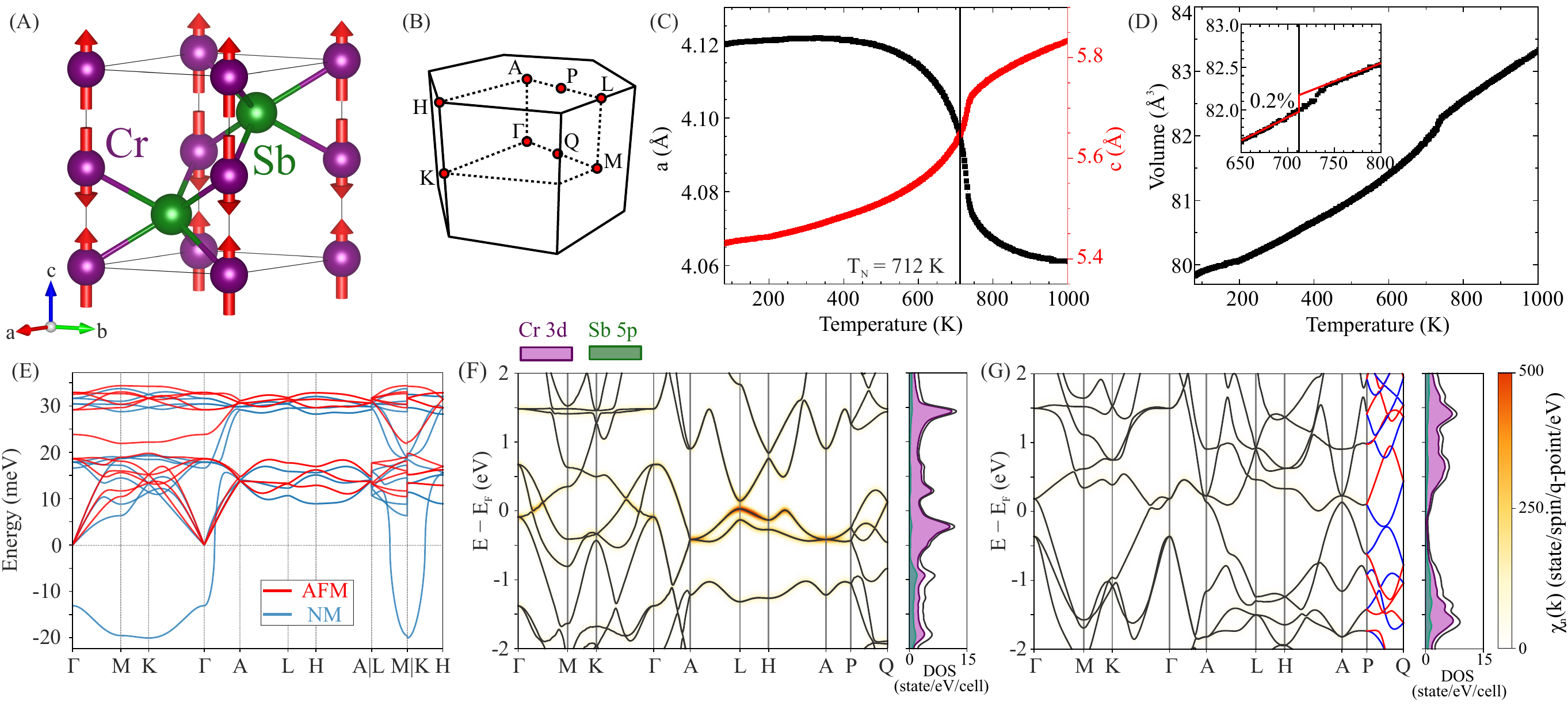}
    \caption{\textbf{Crystal and magnetic structure of CrSb and DFT calculations.} (\textbf{A}) Antiferromagnetic unit cell of CrSb, with alternating collinear magnetic moments on the Cr sites oriented along the \textit{c}-axis. (\textbf{B}) Brillouin zone of CrSb in the space group P6$_3$/\textit{mmc}, showing the high symmetry directions. (\textbf{C-D}) Temperature dependence of the lattice parameters. The temperature dependence of the volume is mostly determined by the \textit{c}-axis expansion with increasing temperature. (\textbf{E}) Harmonic DFPT phonon dispersion of CrSb in the paramagnetic state (blue), overlaid with the magnetic calculations (red). (\textbf{F-G}) Electronic band structure of CrSb obtained from DFT in the non-magnetic (\textbf{F}) and antiferromagnetic phases (\textbf{G}) with the DOS in the side panels. The total DOS phonon DOS (black line) is projected into Cr $3d$ and Sb $5p$ states. The color code in the band structure indicates the strength of the real part of the static band resolved nesting function $\chi_{n}(\mathbf{k})=\sum_m\sum_\mathbf{q}(f_n(\mathbf{k})-f_m(\mathbf{k}+\mathbf{q}))/(\epsilon_n(\mathbf{k})-\epsilon_m(\mathbf{k}+\mathbf{q})+i\delta)$, being $f_n(\mathbf{k})$ the occupation of the state $n$ at wavevector $\mathbf{k}$, $\epsilon_n(\mathbf{k})$ its energy, and $\delta$ a finite infinitesimal real number. The scale bar is common for both plots.
    }
    \label{Fig1}
\end{figure*}

Long-range magnetic order is traditionally rationalized within two complementary paradigms: the ordering of localized atomic moments \cite{Heisenberg_1928} or the instability of nested Fermi surfaces in itinerant systems (SDW) \cite{Gruner_1994a}. By contrast, the emergence of CDWs in metals resists such a unified description \cite{Peierls}. Although often framed in terms of Fermi-surface nesting, this mechanism proves insufficient in higher dimensions (d$>$2), where the geometry of the Fermi surface alone cannot account for the onset or wavevector selection of charge order \cite{Johannes_2008,Weber_TiSe2,Weber_NbSe2,Diego_VSe2,LeTacon_2014,Miao_2019,Miao_2024,Hellmann_2012}. Instead, CDWs frequently arise from a subtle entanglement of electronic, lattice and orbital degrees of freedom, generating a complex hierarchy of coupled instabilities. A large density of states at the Fermi level (E$_\mathrm{F}$) can dramatically enhance electronic correlations and enhance the entanglement between competing/coexisting order parameters. Such conditions naturally arise in lattices with flat bands (FBs), where destructive interference of electronic wavefunctions or restricted hopping suppress the kinetic energy \cite{Checkelsky_2024} and enable interaction-driven phenomena such as fractional quantum Hall states \cite{Wang_2011}, fractional Chern insulators \cite{Regnault_2011,Xie_2021,Kang_2024}, unconventional superconductivity \cite{Tian_2023,Peotta_2015,Wakefield_2023} and exotic magnetism. Magnetic order may therefore coexist — or compete — with hidden translational symmetry breaking, as both instabilities originate from enhanced Coulomb interactions and lattice-mediated attraction, respectively. In the static limit, magnetoelastic coupling \cite{Brown_1965} -- the mutual feedback between spin and lattice -- can modify both magnetic structure and crystal symmetry. Such coupling can amplify small electronic instabilities into macroscopic structural distortions and, conversely, enable magnetic symmetry breaking to reconstruct the phonon spectrum. Beyond its fundamental interest, this spin–lattice entanglement underpins proposals for magnonic and strain-controlled spintronic functionalities \cite{Atulasimha_2011}, where magnetic excitations are manipulated through lattice degrees of freedom.

\begin{figure*}
    \centering
    \includegraphics[width=1.0\linewidth]{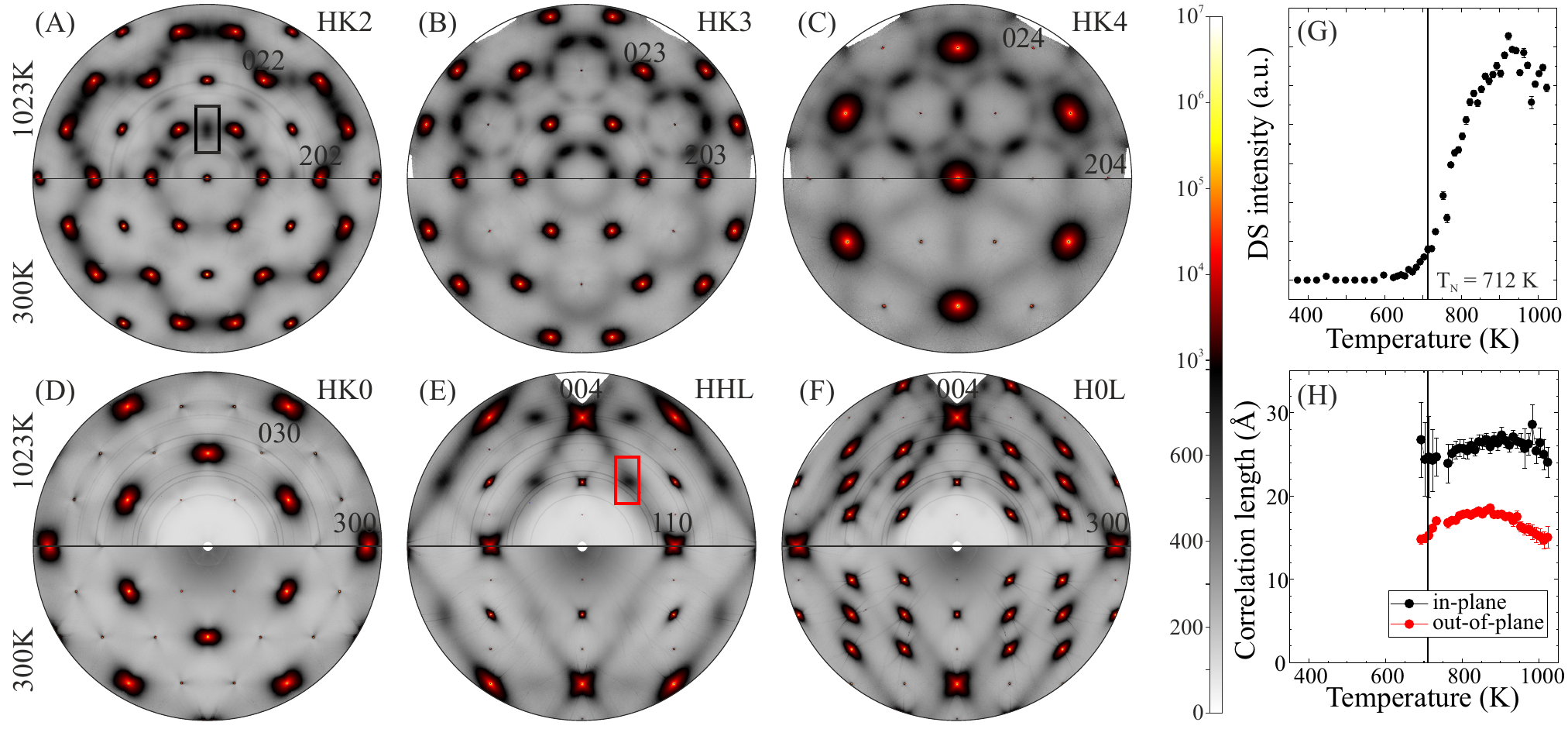}
    \caption{\textbf{Diffuse Scattering results}. (\textbf{A-F}) Comparison of the DS maps measured at 300 K (bottom part) and above $T_\mathrm{N}$ (upper part). Appropriate Laue symmetry was applied to the reconstructed layers to remove detector gaps and improve signal-to-noise ratio. (\textbf{G-H}) Temperature dependence of DS integral intensity and correlation length, respectively, obtained after peak fitting in the regions of interest (ROIs) shown in (\textbf{A}) and (\textbf{E}). The color scheme corresponds to the respective ROIs.  The error bars represent the fit uncertainty.}
    \label{Fig2}
\end{figure*}

In this intertwined regime, lattice and spin excitations cannot be treated independently. Instead, magnons and phonons hybridize and generate composite collective modes \cite{Sukhanov_2025,Koepsell_2019}. While electron–phonon coupling is comparatively better formulated -- with characteristic divergences of the electronic self-energy at q$_\mathrm{CDW}$=2\textit{k}$_\mathrm{F}$ -- the microscopic theory of spin–phonon interactions remains far less developed. The spin–lattice Hamiltonian responsible for magnon–phonon hybridization depends explicitly on both phonon displacements and spin fluctuations, rendering its treatment inherently nonlinear and often anharmonic \cite{Oh_2016}. Although density functional theory (DFT) can capture strain-dependent modifications of exchange interactions and magnetically induced lattice distortions, the dynamical and anharmonic aspects of magnetoelastic hybrid modes remain largely unexplored. Early theoretical work suggested that one-phonon–two-magnon scattering processes can significantly renormalize phonon lifetimes \cite{Silber_1969}, yet systematic experimental validation is still lacking.

Here we report the first direct experimental and theoretical demonstration of flat-electron-phonon coupling in an altermagnetic material \cite{Smejkal_2022,Smejkal_2022a,Li_2025_strain}, which breaks time-reversal symmetry despite hosting collinear antiferromagnetic (AFM) order. In the non-magnetic phase, our harmonic calculations reveal an unstable, nearly dispersionless phonon branch at the $\Gamma$-M-K plane of the Brillouin zone, coexisting with a flat\textit{ish} band pinned at the Fermi level in the \textit{k}$_\mathrm{z}$=$\pi$ plane. This produces a strong and momentum-selective nesting and electron-phonon coupling, enhancing interactions with momenta in-plane with k$_z$=0. Experimentally, we observe short-range charge fluctuations above the Néel temperature, T$>$T$_\mathrm{N}$, providing direct evidence of a flat-band–driven charge instability. At T$_\mathrm{N}$ the acoustic phonon dispersion develops a distinct Kohn-like anomaly, quantitatively captured by stochastic self-consistent harmonic approximation (SSCHA) calculations. Upon entering the AFM state, the diffuse charge signal collapses and the softened mode at M undergoes a large renormalization. This reconstruction reflects the lifting of the FB away from the Fermi level and the onset of strong spin–phonon coupling associated with magnetic symmetry breaking. Our results establish CrSb as the first altermagnet in which flat electronic states are shown to directly control lattice instabilities and magnetoelastic response. More broadly, we demonstrate that altermagnetic symmetry provides a new route to engineer giant, tunable spin–lattice coupling mediated by flat-band physics — opening an unexplored avenue for correlated quantum materials.

CrSb crystallizes in the hexagonal NiAs-type structure (space group \textit{P6$_3$/mmc}, nº 194), in which each Cr atom is octahedrally coordinated by six Sb atoms, forming a lattice with sixfold rotational symmetry. Magnetically, it adopts an A-type AFM configuration: spins align ferromagnetically within the basal plane and stack antiferromagnetically along the \textit{c}-axis, Fig.~\ref{Fig1}(A). Although CrSb shows an overall positive thermal expansion, the Néel transition is accompanied by an abrupt contraction of the \textit{c}-axis of nearly 7$\%$, followed by an anomalous expansion within the \textit{ab}-plane upon further cooling, Fig.~\ref{Fig1}(C). Concomitantly, the unit-cell volume undergoes a drop at T$_\mathrm{N}$ ($\sim$0.2$\%$), providing unambiguous evidence of a sizable magnetoelastic interaction driven by exchange striction, Fig.~\ref{Fig1}(D). This response directly reflects the feedback between spin alignment and interatomic distances at the magnetic transition. The lattice anomalies therefore arise not from an additional structural phase transition, but from a strong coupling between magnetic symmetry breaking and lattice degrees of freedom within the same crystallographic framework.

\begin{figure*}
    \centering
    \includegraphics[width=1.0\linewidth]{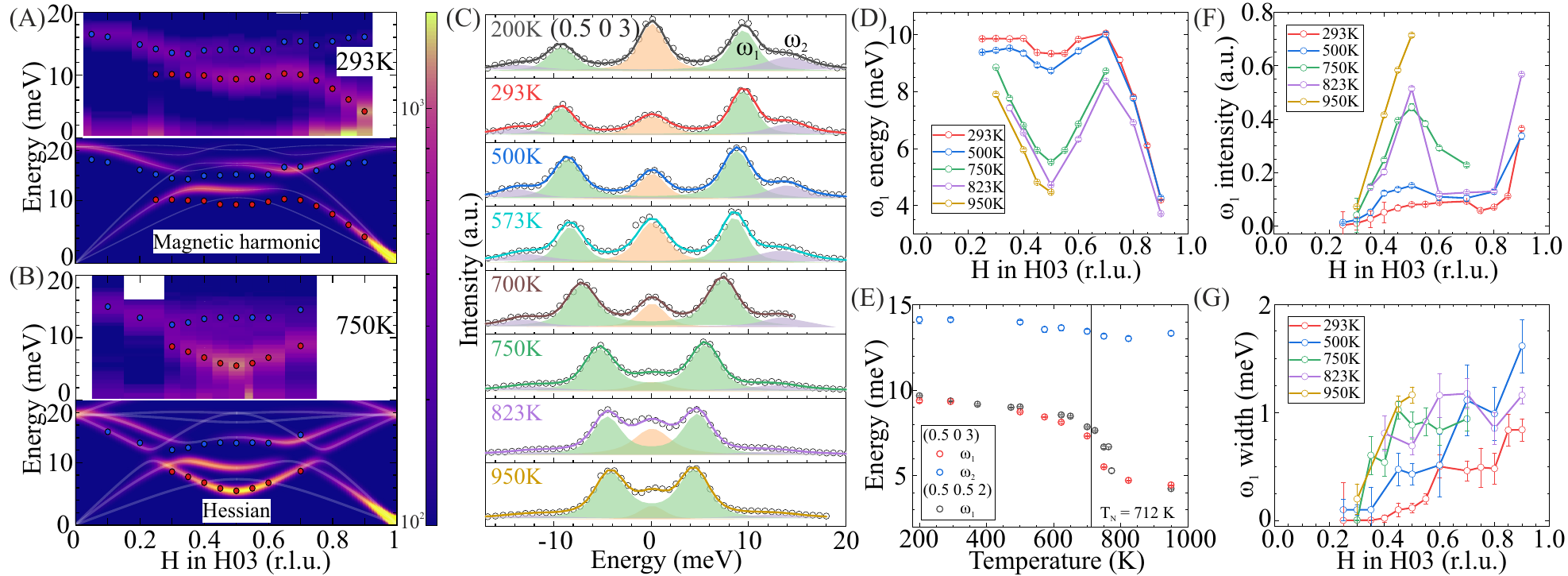}
    \caption{\textbf{Inelastic x-ray scattering and Kohn-like anomaly}. (\textbf{A-B}) Top panels, energy-momentum IXS maps along the H03 direction at 293 K and 750 K, respectively, showing the soft phonon dispersion and enhancement of phonon intensity near the AFM transition temperature. Bottom panels, first-principles calculated IXS spectral functions including the structure-factor $S_n(\mathbf{q}+\mathbf{G})$. An arbitrary linewidth is chosen for visualization. The magnetic and non-magnetic phonons were obtained via DFPT at the harmonic level and considering non-perturbative anharmonicity within the SSCHA at 750 K, respectively. The SSCHA phonons are calculated from the Hessian of the SSCHA free energy (see Ref.~\cite{PhysRevB.96.014111} for more details). (\textbf{C}) Representative IXS spectra at the (0.5 0 3) M point measured between 200 K and 950 K. The elastic line (zero energy loss) and two individual phonon modes are shown as area plots, together with their total fit (solid line) for each spectrum. (\textbf{D}) Fitted $\omega_1$ phonon dispersion as a function of temperature, highlighting the Kohn-like anomaly at the M point at high temperature. (\textbf{E}) Temperature dependence of the $\omega_1$ phonon energy at the M point measured along the H03 and HH2 directions, together with the nearly unchanged higher-energy optic mode $\omega_2$. (\textbf{F-G}) Fitted $\omega_1$ phonon intensity and linewidth, respectively, as a function of momentum transfer at selected temperatures. The error bars represent the fit uncertainty.}
    \label{Fig3}
\end{figure*}

The giant magnetoelastic response is qualitatively captured by the calculation of harmonic phonons within density functional perturbation theory (DFPT)~\cite{RevModPhys.73.515}. The non-magnetic harmonic phonon spectrum, which approximates the paramagnetic state of CrSb above T$_\mathrm{N}$, exhibits an imaginary, nearly dispersionless optical mode in the k$_z$=0 plane, Fig.~\ref{Fig1}(E). From DFPT, this unstable branch corresponds to a coupled vibration involving in-plane displacements of Cr atoms and out-of-plane motions of Sb atoms. Upon imposing AFM order at the DFT level, this soft mode undergoes a dramatic renormalization, stabilizing the lattice and providing microscopic confirmation of the large exchange-striction effect observed experimentally by x-ray diffraction, Fig.~\ref{Fig1}(C–D). We note that due to the large renormalization of the eigenvectors at the AFM phase, the imaginary phonon mode of the non-magnetic calculation cannot be uniquely attributed to a vibrational mode of the AFM phase (see Supplementary Information Fig. S8). Below T$_\mathrm{N}$, the acoustic phonon branches along the $\Gamma$–M direction display additional softening of approximately 5 meV at the M point and $\sim$2 meV at K, further underscoring the strong spin–lattice feedback across the magnetic transition. These changes indicate that magnetic symmetry breaking reshapes not only the unstable optical branch but also the low-energy lattice dynamics. 

Besides, the electronic structure also undergoes a major reconstruction. In the non-magnetic calculation, corresponding to T$>$T$_\mathrm{N}$, DFT reveals hole-like pockets at $\Gamma$ and K, primarily derived from Cr 3\textit{d} states weakly hybridized with Sb orbitals, Fig.~\ref{Fig1}(F). Two nearly flat (weakly dispersive) bands lie close to the Fermi level along the  \textit{k}$_\mathrm{z}$=$\pi$ plane (A-L-H path), generating a peak in the density of states. A third nearly flat band appears approximately 1.4 eV below E$_\mathrm{F}$. On the other hand, the spin-polarized antiferromagnetic calculation reveals a major reorganization of the band structure, Fig.~\ref{Fig1}(G). Kramers degeneracy is lifted out from high symmetry planes of the Brillouin zone, consistent with altermagnetic symmetry, connecting spin-split states by real-space rotational symmetry operations rather than by inversion or translation \cite{Smejkal_2020,Yang_2025,Reimers_2024,Krempaský_2024}. The hole-like pockets at $\Gamma$ and K shift downward by roughly 200 and 500 meV, respectively \cite{Yang_2025,Li_2025}. Most notably, the FBs are completely removed from the vicinity of the Fermi level and pushed to approximately 1.5–2 eV below E$_\mathrm{F}$, leading to a substantial depletion of the density of states at E$_\mathrm{F}$ in the altermagnetic phase given their more dispersive character, side panel of Fig.~\ref{Fig1}(G). This dramatic electronic reconstruction directly links magnetic ordering to the stabilization of the lattice: the lifting of flat electronic states from the Fermi level suppresses the electronic instability that drives the soft phonon mode in Fig.~\ref{Fig1}(D). 

\begin{figure*}
    \centering
    \includegraphics[width=\linewidth]{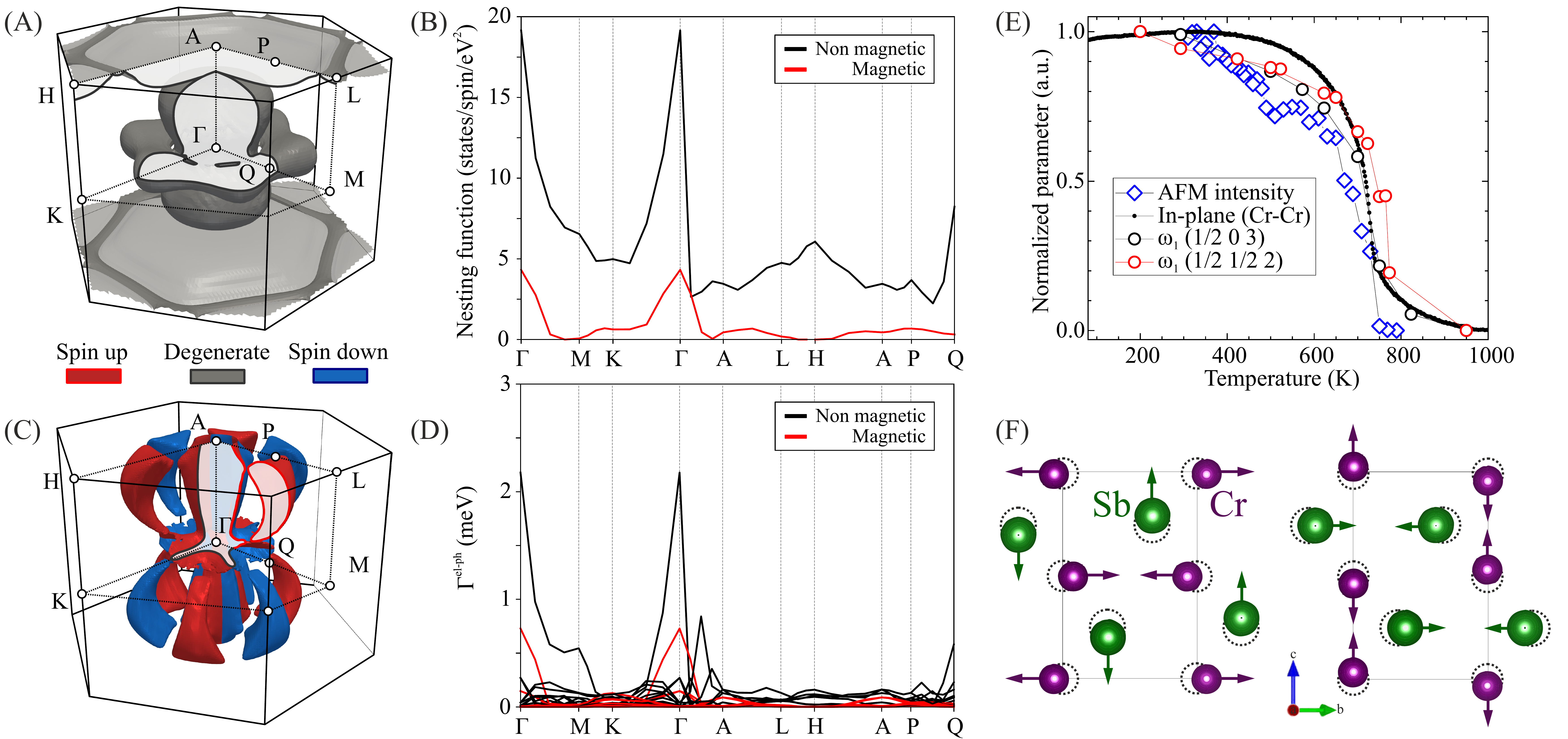}
    \caption{\textbf{Fermiology and linear response electron-phonon coupling in CrSb.} (\textbf{A}) Fermi-surface of the non-magnetic phase. (\textbf{B}) Calculated nesting function $\chi_0''(\mathbf{q})=\sum_{nm\mathbf{k}}\delta(\epsilon_n(\mathbf{k})-E_F)\delta(\epsilon_m(\mathbf{k}+\mathbf{q})-E_F)$ for the non-magnetic (black) and magnetic (red) phases. (\textbf{C}) Spin-resolved Fermi surface of the magnetic phase. Red and blue tones denote spin-up and spin-down bands, respectively. (\textbf{D}) Linear-response electron–phonon linewidths for the non-magnetic (black) and magnetic (red) phases. (\textbf{E}) Temperature dependence of the AFM neutron scattering intensity (blue diamonds, adapted from \cite{singh2025chiral}), the nearest-neighbour in-plane Cr--Cr distance (black filled circles), and the phonon intensity of the $\omega_{1}$  mode (open circles). (\textbf{F}) Candidate lattice distortions of the non-magnetic phase involving in-plane and out-of-plane displacements of Cr and Sb atoms.}
    \label{Fig4}
\end{figure*}

To experimentally prove the theoretically predicted lattice instability associated with the imaginary flat phonon mode -- and its coupling to magnetism -- we carried out diffuse (DS) and inelastic x-ray scattering (IXS) measurements. At high temperatures (T$>$1000 K), the DS reveals broad structured intensity between Bragg reflections at integer L values, with an enhancement at the M point of the Brillouin zone, \textit{q}*=($\frac{1}{2},\ 0$), top panels of Fig.~\ref{Fig2}(A-C). In addition, the diffuse signal is absent in the HK0 plane, Fig.~\ref{Fig2}(D), demonstrating that the correlated disorder is dominated by phonon fluctuations. In the HKL planes (L$\neq$ 0), the diffuse intensity exhibits strong directionality, indicative of anisotropic electron–phonon coupling, likely involving in-plane phonon displacements. Remarkably, the structured diffuse intensity weakens upon cooling below T$_\mathrm{N}$, and becomes drastically reduced at room temperature. At 300 K, the diffuse hexagonal clouds pinned at M (HK3 map) evolve into nearly uncorrelated anisotropic in-plane rings of intensity, whereas the HK4 plane displays diffuse streaks connecting Bragg points, bottom panel Fig.~\ref{Fig2}(A-C). Complementary HHL and H0L maps, Fig.~\ref{Fig2}(E-F) reveal a temperature-driven crossover: at high temperature, the spectral weight concentrates near half-integer momenta, while upon cooling the diffuse streaks extend along the K–A direction, signaling the gradual suppression of short-range charge correlations. Fig.~\ref{Fig2}(G) summarizes the temperature evolution of the diffuse intensity within the region of interest highlighted in Fig.~\ref{Fig2}(A). The DS intensity increases markedly between 700 K and 1000 K, pointing to a broad regime of competing interactions in which short-range charge correlations progressively develop as AFM order is suppressed. Despite this enhancement in intensity, the associated correlation lengths remain short and nearly temperature-independent, with characteristic values of approximately 25 \r{A} within the \textit{ab}-plane and 15 \r{A} along the \textit{c}-axis, Fig.~\ref{Fig2}(H). This behavior indicates an unconventional thermally driven transition characterized not by long-range symmetry breaking, but by the emergence of robust short-range correlations that effectively lower the local symmetry of the lattice. The persistence of limited correlation lengths, together with the strong temperature-dependent spectral weight redistribution of the DS, highlights the cooperative role of electron–phonon and spin–phonon interactions in stabilizing a fluctuating charge state that competes directly with long-range magnetic order. It is worth noting that in binary compounds the transition from the hexagonal to the orthorhombic structure has been observed under high pressure \cite{PhysRevB.10.1248}. This transformation typically involves in-plane displacements of the magnetic atoms and out-of-plane buckling of the non-magnetic sublattice, Fig.~\ref{Fig4}(F). A related behavior was recently proposed theoretically in MnAs \cite{PhysRevLett.104.147205, PhysRevB.83.054108}, where a structural phase transition from the high-symmetry hexagonal phase to the low-symmetry orthorhombic phase was shown to be mediated by a phonon softening.

The analysis above demonstrates that the short-range charge fluctuations observed at the M point of the BZ originate from strong electron–(spin)–phonon coupling associated with a soft phonon mode, which is directly accessible via IXS. The IXS spectra were collected near the reciprocal lattice vector $\overrightarrow{\textit{G}}$=(0\ 0\ 3), mapping the phonon dispersion as a function of momentum transfer \textit{Q}. Fig.~\ref{Fig3}(A-B) show representative IXS intensity maps measured at 293 K and above T$_\mathrm{N}$. The experimental spectra are in excellent agreement with the phonon frequencies obtained including anharmonic effects within the stochastic self-consistent harmonic approximation (SSCHA)~\cite{Errea_2014,Monacelli_2021,PhysRevB.96.014111}, confirming the reliability of the theoretical description when lattice anharmonicity is included in the calculations. At the critical wavevector q$^*$=($\frac{1}{2}\ 0\ 3$), the inelastic scans reveal two prominent phonon excitations with energies of approximately 10 meV and 14 meV, denoted as $\omega_1$ and $\omega_2$ in Fig.~\ref{Fig3}(C). The corresponding lattice vibrations of the lowest energy mode involves hybridized atomic displacements combining in-plane motions of Cr atoms with out-of-plane motions of Sb atoms. This mixed displacement character reflects the eigenvector composition of the soft mode, which cannot be associated with purely Cr or Sb vibrations but instead represents a collective lattice excitation of the coupled Cr–Sb sublattice. The observation of these finite-energy modes thus indicates that magnetic ordering stabilizes the lattice by lifting the imaginary phonon anomaly present in the non-magnetic phase, providing direct experimental evidence for the magnetic quenching of the flat-band–driven lattice softening.

In the momentum range 0.5$<$H$<$1 r.l.u., the dispersion of the $\omega_1$ mode at 293 K is primarily governed by the acoustic branch, which deviates from a simple sinusoidal dispersion near the M point, where a small but resolvable dip is observed beyond the experimental uncertainty (Fig.~\ref{Fig3}D). At lower momentum transfers, 0.0$<$H$<$0.5 r.l.u., the IXS spectra exhibit a mixed character, with optical-branch contributions dominating the range 0.0$<$H$<$0.25 and acoustic-branch weight becoming more prominent between 0.25$<$H$<$0.5. To extract quantitative information on phonon energies and linewidths, the IXS spectra were fitted using damped harmonic oscillator line shapes convoluted with the instrumental resolution function ($\sim$3 meV) (see Fig.~\ref{Fig3}(C) and Supplementary Information S1-S3). The $\omega_2$ branch remains nearly dispersionless across momentum space and exhibits only a softening of approximately 1 meV at the M point between room temperature and 950 K, consistent with anharmonic temperature effects predicted by the SSCHA calculations. In contrast, the $\omega_1$ mode shows a pronounced temperature dependence as the system approaches T$_\mathrm{N}$. The dip at the M point becomes progressively sharper, accompanied by a phonon softening of approximately 6 meV at T$_\mathrm{N}$, Fig.~\ref{Fig3}(D). This behavior agrees well with the giant exchange-striction–induced phonon renormalization obtained from magnetic harmonic DFPT simulations (Fig.~\ref{Fig1}(E)). Importantly, the strong coupling between the $\omega_1$ mode and AFM fluctuations is highly localized in momentum space around q$^*$=($\frac{1}{2}\ 0$). Upon approaching T$_\mathrm{N}$, phonon fluctuations become increasingly suppressed in this region, leading to the formation of a Kohn-like anomaly that competes with magnetic symmetry breaking. Above T$_\mathrm{N}$, the $\omega_1$ frequency stabilizes near $\sim$4 meV, again in excellent agreement with SSCHA predictions in Fig.~\ref{Fig3}(B), which address that the high-temperature non-magnetic phase is stabilized thanks to lattice anharmonicity. The observed phonon renormalization of approximately 6 meV represents one of the largest spin–phonon coupling effects reported, exceeding the values typically observed in multiferroic oxides \cite{Calder_2015,Rudolf_2007,Kamba_2014,Sakai_2011,Garcia_2012}. Furthermore, the integrated intensity of the $\omega_1$ mode increases with temperature, Fig.~\ref{Fig3}(F), suggesting that this excitation contributes dominantly to the DS shown in Fig.~\ref{Fig2}(G). Although the temperature dependence of the soft mode follows a continuous evolution consistent with phonon coupling to the magnetic order parameter, the extracted critical exponent from fitting the temperature-dependent frequency $\omega_1$(\textit{T$^*$}), where \textit{T$^*$} is the reduced temperature, returns a value of $\beta$=0.12$\pm$0.01, which is smaller than the AFM order parameter \cite{singh2025chiral}. This indicates that the lattice response is more strongly fluctuating than the magnetic order itself. 

We now discuss the microscopic mechanism underlying the hidden charge fluctuations, and examine the relative roles of electron-phonon interaction and Fermi-surface nesting. Fig.~\ref{Fig4}(A) shows the DFT calculated three-dimensional Fermi surface of CrSb in the high-temperature phase (T$>$T$_\mathrm{N}$). The FS sheet forms a sixfold hexagonally anisotropic contour at \textit{k}$_z$=0, with vertices oriented toward the K point and exhibiting an hourglass-like dispersion along the $\Gamma$–A direction. In the \textit{k}$_z$=$\pi$ plane, the Fermi surface develops a weakly dispersive, nearly flat contour in the \textit{k}$_x$-\textit{k}$_y$ projection. This flat electronic manifold is responsible for the enhanced density of states near the Fermi level and provides multiple momentum channels for potential nesting-driven charge instabilities. To quantify nesting tendencies, we evaluate the nesting function derived from the imaginary part of the bare electronic susceptibility, $\chi_0^{\prime\prime}$(q). The calculation reveals enhanced susceptibility in the $\Gamma$-M-K plane, directly associated with the flat electronic plane, Fig.~\ref{Fig4}(B). Additional large contributions appear around H-L points, consistent with transitions from states around the $\Gamma$-M-K plane to flat region in the \textit{k}$_z$=0 plane. In the AFM phase (T$<$ T$_\mathrm{N}$), the Fermi surface reconstructs into a closed three-dimensional ellipsoidal pocket with spin-split band branches characteristic of altermagnetic symmetry, Fig.~\ref{Fig4}(C). Viewed along the crystallographic \textit{c}-axis, the low-temperature Fermi surface exhibits a characteristic \textit{g}-wave–like symmetry, with nodal planes enforcing spin degeneracy through symmetry constraints. Notably, the flat electronic contour present at \textit{k}$_z$=$\pi$ in the high-temperature phase is completely removed in the magnetic state. As a consequence, the maxima in $\chi_0^{\prime\prime}$(q) collapse, and all nesting-driven instability channels are suppressed, red curve in Fig.~\ref{Fig4}(B). Although a perfectly nested flat plane at \textit{k}$_z$=$\pi$ would favor instability near the H point, the divergence of the imaginary part of the susceptibility alone does not guarantee CDW formation \cite{Johannes_2008}. Instead, CDW stability is determined by the combined contribution of the real and imaginary parts of the susceptibility and the momentum-dependent electron–phonon matrix elements.

Consistent with this picture, the calculated electron–phonon linewidth exhibits strong momentum localization, with strong coupling at the M point and nearly vanishing interaction strength at the H and Q points, Fig.~\ref{Fig4}(D). This behavior indicates that the electron–phonon matrix elements are highly mode- and momentum-selective, effectively confining charge fluctuations to the M point. In the magnetic phase, the electron–phonon coupling is significantly reduced, implying that the energy scale associated with altermagnetic ordering exceeds the relevant electron–phonon interaction strength. The resulting \textit{failed} CDW scenario reflects competition between magnetic symmetry breaking and lattice instability. Accordingly, the giant phonon renormalization observed below T$_\mathrm{N}$ originates from exchange-striction–mediated coupling between in-plane Cr ions --direct Cr--Cr interactions and double exchange via Cr--Sb--Cr bonds-- stabilizing the high-symmetry hexagonal structure in the magnetically ordered phase. The temperature evolution of the Cr–Cr bond distance provides direct structural evidence of this magnetoelastic interaction, Fig.~\ref{Fig4}(E). The calculated phonon energy shift agrees quantitatively with experiment and is driven primarily by the ferromagnetic Cr–Cr exchange interactions within the hexagonal lattice, establishing exchange striction as the microscopic origin of the giant magnetoelastic response.

The competition between charge and spin order revealed in CrSb places this altermagnet within the broader landscape of quantum materials where intertwined orders govern the emergent ground state. In high-temperature cuprates \cite{Keimer_2015} and the recently discovered superconducting nickelates \cite{Li_2019}, superconductivity develops in close proximity to competing spin and charge density wave instabilities \cite{Ghiringhelli_2012,Chang_2012,Blanco_2013,Tranquada_1995}, highlighting how small changes in electronic correlations can reorganize collective order parameters. Similarly, moiré materials such as twisted graphene systems and kagome metals exhibit flat electronic bands that amplify interactions, giving rise to correlated insulating states, superconductivity, and symmetry-breaking orders that compete or coexist \cite{Cao_2018,Cao_2018a,Balents_2020,Chen_2020,Ortiz_2019,Zhao_2021,Wilson_2024,Kang_2023,Teng_2022}. In heavy-fermion compounds, the competition between magnetic order, superconductivity and Kondo screening leads to quantum criticality and emergent collective phases \cite{Doniach_1977, Steglich_1979, Paschen_2004, Mathur_1998}. In this context, CrSb provides a particularly transparent platform: its symmetry-protected altermagnetic state and flat-band electronic structure generate a direct competition between charge and spin order that strongly reconstructs the lattice dynamics. As in these other correlated systems, the delicate balance between intertwined order parameters determines the macroscopic ground state, underscoring the universal role of competing interactions in strongly correlated quantum materials.


In conclusion, we show that the magnetic transition in CrSb is exceptional: the system tends to lower its lattice symmetry upon heating due to the development of short-range structural correlations, even if no structural transition occurs before melting. Our results therefore provide direct experimental evidence, supported by theoretical calculations, for strong spin-phonon coupling driving the magnetostructural transition in CrSb. CrSb, therefore, provides the first direct demonstration that altermagnetic order quenches a flat-band–driven lattice instability through momentum-selective electron–phonon coupling. The competition between nesting, strong electron-phonon coupling at M, and exchange striction suppresses a putative CDW, stabilizing the magnetic phase and establishing a microscopic link between flat bands, spin–phonon coupling, and giant magnetoelasticity.

\section*{Methods}
The CrSb single crystals were grown by the chemical vapor transport (CVT) method. A stoichiometric ratio of chromium and antimony powders, together with iodine of 2.5 mg/ml as the transport agent, were mixed and sealed in an evacuated quartz ampoule. The ampoule was slowly heated and finally exposed to a temperature gradient of 925°C to 900°C where the CVT preceded for one week, then naturally cooled down to room temperature. CrSb crystals in size of 5-10 mm with regular shapes and shiny surfaces were obtained.

X-ray single-crystal diffraction measurements were carried out at the BM01 beamline (Swiss--Norwegian Beamlines, SNBL) of the European Synchrotron Radiation Facility (ESRF) using an incident energy of 17.2~keV and a Dectris PILATUS~2M area detector~\cite{ dyadkin2016new}. Low-temperature measurements in the range of 80--400~K were performed using a Cryostream N$_2$ blower (Oxford Cryosystems), while a solid-state furnace was used for high-temperature measurements up to 1000~K. Diffraction images were collected with a 1$^\circ$ step in angular rotation over a full 360$^\circ$ range. The raw data were processed using the \textit{SNBL Toolbox}, and sequential refinements at different temperatures were performed with \textit{CrysAlisPro} software.

Single-crystal diffuse scattering measurements were conducted at the side-station of the ID28 beamline (ESRF) using an incident energy of 17.8~keV and a Dectris PILATUS3~1M area detector. Temperature control was achieved with N$_2$ heatblower. High-flux data, emphasizing low-intensity diffuse features between oversaturated Bragg reflections, were collected with an angular step of 0.25$^\circ$ over a full 360$^\circ$ range. The orientation matrix was refined using \textit{CrysAlis} package and subsequently employed for reciprocal-space reconstructions with the beamline-specific \textit{ProjectN} software. The resulting reciprocal-space maps were visualized in the \textit{Albula} (Dectris).

Inelastic X-ray scattering (IXS) experiments were performed at the ID28 beamline (ESRF) using a Si(999) backscattering monochromator (17.8~keV). The overall energy resolution was determined from measurements of a standard plexiglas sample at low temperature, where fitting the elastic line with a pseudo-Voigt profile yields a full width at half maximum of $\Delta E = 3$~meV. Measurements were conducted in transmission geometry at constant temperatures, controlled using a Cryostream N$_2$ blower (Oxford Cryosystems) and N$_2$ heatblower. The components $(H~K~L)$ of the scattering vector $\mathbf{Q}$ are expressed in reciprocal lattice units (r.l.u.) as $\mathbf{Q} = H\mathbf{a}^* + K\mathbf{b}^* + L\mathbf{c}^*$, where $\mathbf{a}^*$, $\mathbf{b}^*$, and $\mathbf{c}^*$ are the reciprocal lattice vectors. All IXS spectra were fitted using a single damped harmonic oscillator (DHO) model, implemented in beamline-specific software \textit{FIT28}. The elastic line and phonon excitations were extracted by deconvolving the experimental spectra with the instrumental resolution function. The energy position, linewidth, and intensity of each phonon mode were used as fitting parameters. The Stokes and anti-Stokes components were corrected for the Bose--Einstein thermal population factor.

DFT calculations were performed using the Quantum Espresso package~\cite{Giannozzi_2009,Giannozzi_2017,Giannozzi_2020}, with the Perdew-Burke-Ernzerhof parametrization of the generalized gradient approximation~\cite{Perdew_1996}. An 80/800 Ry cutoffs were employed for wavefunctions/density, where the Brillouin zone was sampled by a $12\times12\times12$ regular $\mathbf{k}$-mesh with a 0.005 Ry Methfessel-Paxton smearing~\cite{Methfessel_1989}. For the magnetic phase, collinear magnetism was considered. For visualization of electronic bands, and Fermi surfaces, a non self-consistent calculation in a regular $36\times36\times36$ $\mathbf{k}$-grid was performed.

Harmonic phonons were computed by density functional perturbation theory (DFPT)~\cite{RevModPhys.73.515} implemented in the Quantum Espresso package~\cite{Giannozzi_2009,Giannozzi_2017}. A $2\times2\times2$ $\mathbf{q}$-grid was considered, which captures the high-symmetry wavevectors where the relevant physics occur. In order to prove the stability of the non-magnetic phase above the $T_C$, its temperature dependent anharmonic phonon spectra were computed employing the stochastic self-consistent harmonic approximation (SSCHA)~\cite{Errea_2014,PhysRevB.96.014111,Monacelli_2021}. SSCHA calculation of the ionic distribution density, free energy and anharmonic phonons obtained from the Hessian od the free energy were computed in a $2\times2\times2$ $\mathbf{q}$-grid, employing DFT self-consistent calculations with the above mentioned setup to compute energies and forces, needed to perform the SSCHA variational minimization. The calculation of the dynamical structure factor is presented in the Supplementary Information.


\section*{Author Contributions}
S.B-C conceived and managed the project. C.L. and Z.L. synthesized and characterized the samples. A.K. and A.B. carried out the DS and IXS measurements. A.K., C-Y.Lim and D.C performed the diffraction experiments. A.K. analyzed the experimental data. M.A., F.B., M.G.V and I.E. performed the DFT and phonon calculations. S.B-C wrote the manuscript with input from all co-authors. Samples request: Cong Li, conli@kth.se

\section*{Acknowledgments}
We thank Victor Pardo, Rafael Fernandes, Matthieu LeTacon. Eduardo da Silva-Neto, A. Frano and Tianping Ying for fruitful discussions and critical reading of the manuscript. A.K. thanks the Basque government for financial support through the project PIBA-2023-1-0051. S.B-C. thanks the MINECO of Spain, projects PID2021-122609NB-C21 and PID2024-161503NB-C21. C.-Y.L. is supported by the IKUR Strategy under the collaboration agreement between Ikerbasque Foundation and DIPC on behalf of the Department of Education of the Basque Government. M.G.V received financial support from the Canada Excellence Research Chairs Program for Topological Quantum Matter.  M.G.V and F.B. thank support to the Spanish Ministerio de Ciencia e Innovacion grant PID2022-142008NB-I00 and the Ministry for Digital Transformation and of Civil Service of the Spanish Government through the QUANTUM ENIA project call - Quantum Spain project, and by the European Union through the Recovery, Transformation and Resilience Plan - NextGenerationEU within the framework of the Digital Spain 2026 Agenda. I.E. also acknowledges financial support from the Department of Education, Universities and Research of the Eusko Jaurlaritza, and the University of the Basque Country UPV/EHU (Grant No. IT1527-22) and the Spanish Ministerio de Ciencia e Innovación (Grant No. PID2022-142861NA-I00). 

\section*{Competing Interests}
The authors declare no competing interests.

\bibliography{CrSb}

\end{document}


\def\qq{\mathbf{q}}
\newcommand{\red}[1]{\textcolor{red}{#1}}
\newcommand{\blue}[1]{\textcolor{blue}{#1}}
\newcommand{\ion}[1]{\textcolor{purple}{\textbf{(Ion: #1)}}}

\title{\textit{Supplementary Materials}: Flat band driven competing charge and spin instabilities in the altermagnet CrSb}

\author{A. Korshunov}
\thanks{These authors contributed equally.}
\affiliation{Donostia International Physics Center (DIPC), San Sebastián, Spain}

\author{M. Alkorta}
\thanks{These authors contributed equally.}
\affiliation{Centro de Física de Materiales (CFM-MPC), CSIC-UPV/EHU, San Sebastián, Spain}
\affiliation{Fisika Aplikatua Saila, Gipuzkoako Ingeniaritza Eskola, University of the Basque Country (UPV/EHU), San Sebastián, Spain}

\author{C.-Y. Lim}
\affiliation{Donostia International Physics Center (DIPC), San Sebastián, Spain}

\author{F. Ballester}
\affiliation{Donostia International Physics Center (DIPC), San Sebastián, Spain}
\affiliation{Fisika Aplikatua Saila, Gipuzkoako Ingeniaritza Eskola, University of the Basque Country (UPV/EHU), San Sebastián, Spain}

\author{Cong Li}
\affiliation{Department of Applied Physics, KTH Royal Institute of Technology, Stockholm 11419, Sweden}
\affiliation{Beijing National Laboratory for Condensed Matter Physics, Institute of Physics, Chinese Academy of Sciences, Beijing 100190, China} 

\author{Zhilin Li}
\affiliation{Beijing National Laboratory for Condensed Matter Physics, Institute of Physics, Chinese Academy of Sciences, Beijing 100190, China}

\author{D. Chernyshov}
\affiliation{Swiss-Norwegian BeamLines at European Synchrotron Radiation Facility, BP 220, F-38043 Grenoble Cedex, France}

\author{A. Bosak}
\affiliation{European Synchrotron Radiation Facility (ESRF), BP 220, F-38043 Grenoble Cedex, France}

\author{M. G. Vergniory}
\affiliation{Donostia International Physics Center (DIPC), San Sebastián, Spain}
\affiliation{Département de Physique et Institut Quantique, Université de Sherbrooke, Sherbrooke, Québec, Canada}

\author{Ion Errea}
\email{ion.errea@ehu.eus}
\affiliation{Donostia International Physics Center (DIPC), San Sebastián, Spain}
\affiliation{Centro de Física de Materiales (CFM-MPC), CSIC-UPV/EHU, San Sebastián, Spain}
\affiliation{Fisika Aplikatua Saila, Gipuzkoako Ingeniaritza Eskola, University of the Basque Country (UPV/EHU), San Sebastián, Spain}

\author{S. Blanco-Canosa}
\email{sblanco@dipc.org}
\affiliation{Donostia International Physics Center (DIPC), San Sebastián, Spain}
\affiliation{IKERBASQUE, Basque Foundation for Science, 48013 Bilbao, Spain}

\maketitle

\section{Inelastic X-ray scattering (IXS)}

\begin{figure*}
    \centering
    \includegraphics[width=\linewidth]{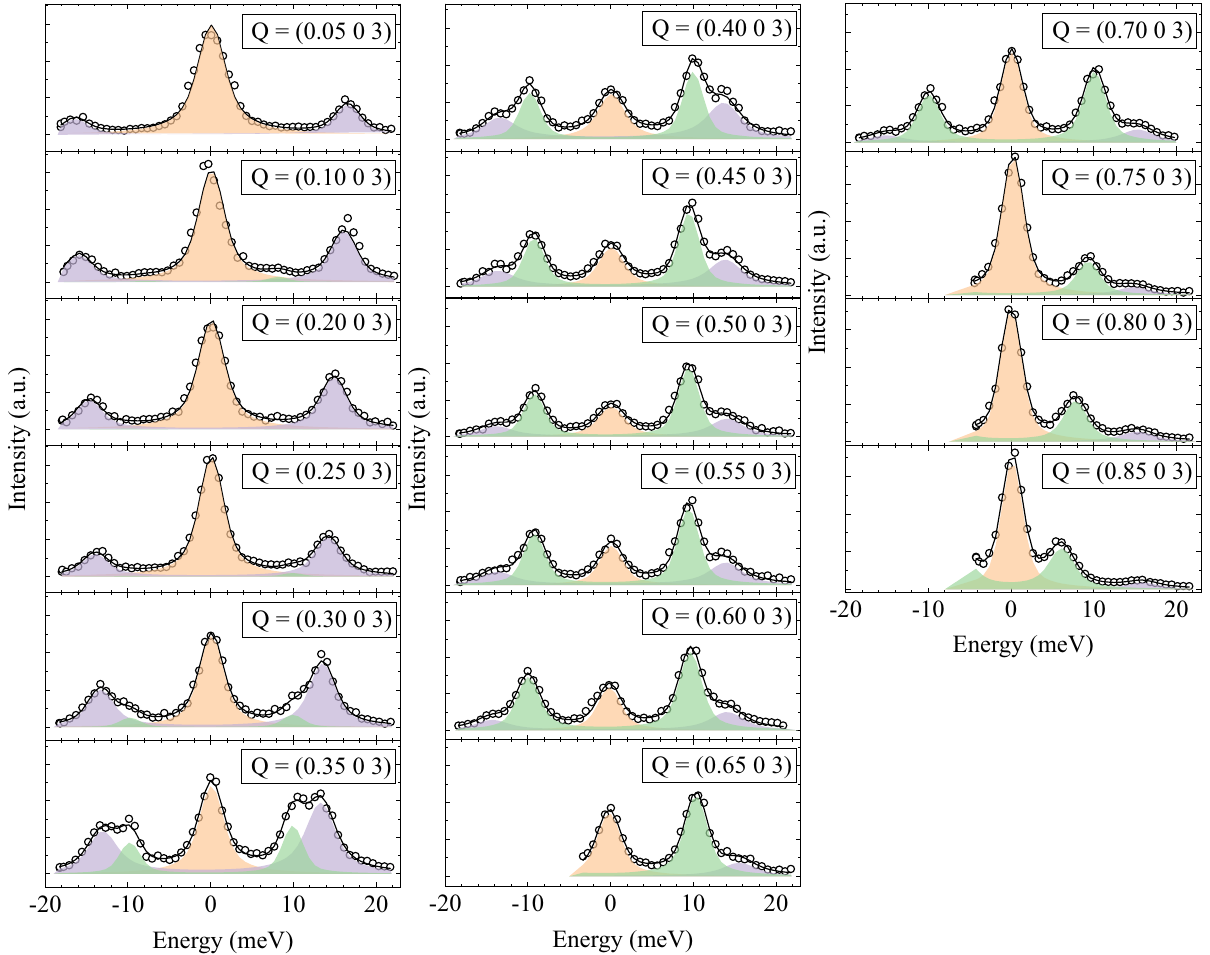}
    \caption{Fitted IXS spectra measured at RT along the H03 direction. Individual Q points are indicated in the panels. The elastic line (zero energy loss) and two individual phonon modes are shown as area plots together with their total fit (solid line) for each spectrum. The intensity scale is not the same for all panels.}
    \label{SI_IXS1}
\end{figure*}

\begin{figure*}
    \centering
    \includegraphics[width=\linewidth]{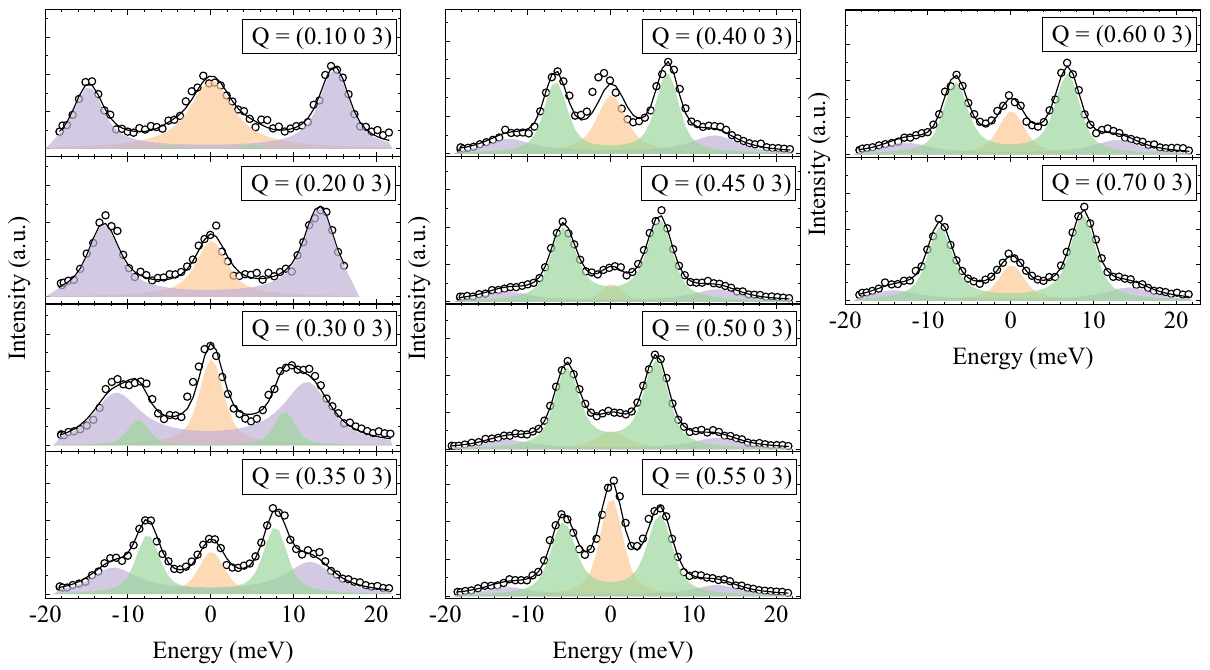}
    \caption{Fitted IXS spectra measured at 750K along the H03 direction. Individual Q points are indicated in the panels. The elastic line (zero energy loss) and two individual phonon modes are shown as area plots together with their total fit (solid line) for each spectrum. The intensity scale is not the same for all panels. The apparent shift of the elastic line in the spectrum measured at $(0.4,0,3)$ originates from a spurious contribution to the elastic signal. The inelastic energy scale follows the same linear dependence on the main monochromator temperature as for the other spectra.}
    \label{SI_IXS2}
\end{figure*}

\begin{figure*}
    \centering
    \includegraphics[width=\linewidth]{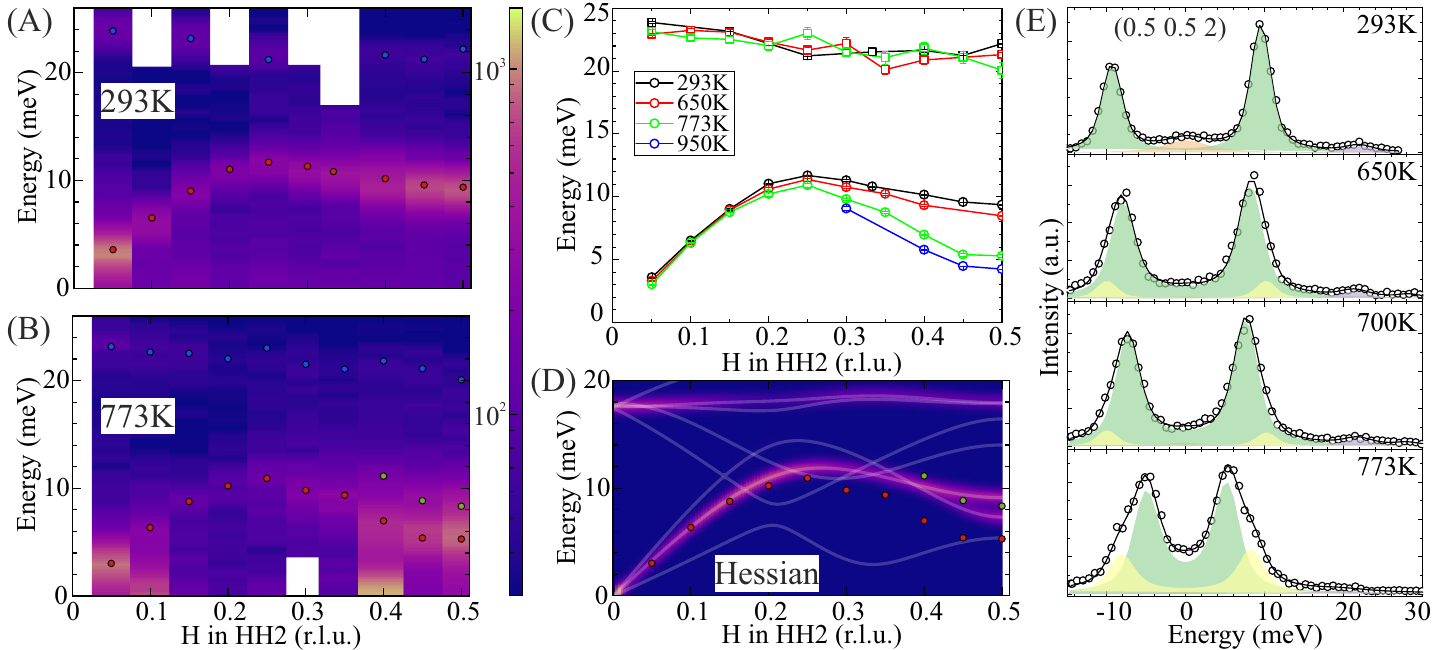}
    \caption{(A,B) Energy-momentum IXS maps along the HH2 direction at 293 K and 773 K, respectively, showing the soft phonon dispersion and enhancement of phonon intensity above the AFM transition temperature. (C) Fitted phonon dispersion along HH2 as a function of temperature, highlighting the Kohn-like anomaly at the M point at high temperature. (D) First-principles calculated IXS spectral function including the structure-factor $S_n(\mathbf{q}+\mathbf{G})$. An arbitrary linewidth is chosen for visualization. The SSCHA phonons are calculated from the Hessian of the SSCHA free energy at 750K.
    (E) Representative IXS spectra at the (0.5 0.5 2) M point measured between 293 K and 773 K. The elastic line (zero energy loss) and three individual phonon modes are shown as area plots, together with their total fit (solid line) for each spectrum.  }
    \label{SI_IXS3}
\end{figure*}

In the following Supplementary Figs.~\ref{SI_IXS1}-\ref{SI_IXS3}, we present a detailed description of the IXS measurements as a function of temperature and momentum. The fitting results at the room-temperature (RT) and 750~K (corresponding colormaps are shown in Fig.~3(A,B) of the main text) are presented in Figs.~\ref{SI_IXS1} and \ref{SI_IXS2}, respectively. All measured momentum transfers are specified in reciprocal lattice units (r.l.u.) and given in the HKL notation.

Along the H03 direction, an acoustic-like phonon mode $\omega_1$ dominates the spectra for $H > 0.35$~r.l.u. In the vicinity of $H \approx 0.35$~r.l.u., a transfer of spectral weight occurs between the acoustic mode $\omega_1$ and an optic mode $\omega_2$, with the optic mode becoming dominant at smaller $H$. At RT, well below the Néel temperature $T_N = 712$~K, the acoustic branch exhibits a pronounced deviation from a simple sine-like dispersion at M point, in agreement with density-functional-theory (DFT) calculations. Upon heating, the $\omega_1$ mode significantly softens, and the dip at the M point becomes more pronounced, as shown in Fig.~3(A-C) of the main text.

An equivalent behavior is observed at the symmetry-related M point in the HK2 reciprocal-space plane, where strong diffuse scattering (DS) was also detected, Fig.~\ref{SI_IXS3}(A-C). In this direction, a small splitting of the acoustic mode is also resolved, originating from the lifting of acoustic phonon degeneracy, as illustrated in the theoretically calculated dispersion map Fig.~\ref{SI_IXS3}(D). The temperature dependence of the M-point spectrum, measured at $(\tfrac{1}{2},\tfrac{1}{2},2)$ reveals a pronounced enhancement of this splitting upon approaching $T_N$, as shown in Fig.~\ref{SI_IXS3}(E).

According to the DFT calculations shown in Fig.~1E of the main text, the optic phonon mode at approximately 25~meV exhibits the strongest renormalization driven by spin--phonon coupling. Experimentally, however, no clear softening of this mode is observed, Fig.~\ref{SI_IXS3}(C). Instead, the IXS data reveal a pronounced increase of its damping with temperature, manifested by a substantial linewidth broadening, as shown in Fig.~\ref{SI_IXS3}(E). This behavior indicates a dynamical instability of the optic mode that does not result in a coherent phonon condensation. Rather, the instability is transferred to lower energies through strong hybridization and spectral-weight redistribution between optic and acoustic phonon branches. As a consequence, the dominant lattice instability at the M point is represented as a softening of the $\omega_1$ phonon mode.

\section{Symmetry analysis of the dynamic phonon instability}

At ambient conditions, CrSb crystallizes in the hexagonal P6$_3$/mmc (194) space group, with two crystallographically independent atomic sites (Cr at 2a and Sb at 2c), resulting in a total of four atoms per unit cell. The presence of a $6_3$ screw axis, combined with mirror symmetry, implies a $c$-glide plane, manifested in the relative shift of the Sb atomic position along the $c$ direction. The extinction conditions of the $P6_3/mmc$ space group lead to systematic absences of reflections with odd L (L = 2n+1) along the 00L direction, as well as in the HHL plane. Despite these extinction rules, high-intensity diffuse measurements reveal weak, temperature-independent peaks at symmetry-forbidden positions, both in the HK-planes with odd L and in HHL planes. Their intensity is at least two orders of magnitude lower than that of the main Bragg reflections and can therefore be overlooked in standard structural diffraction. Their presence is attributed to structural defects associated with a local violation of the 6$_3$ screw symmetry, effectively leading to a doubling of the unit cell along the $c$ direction, most likely originating from stacking faults. Apart from these weak defect-related features, diffraction data collected over the entire temperature range indicate high crystal quality, with no evidence for macroscopic symmetry lowering: all Bragg reflections remain sharp and show no detectable splitting.

\begin{figure}
    \centering
    \includegraphics[width=\linewidth]{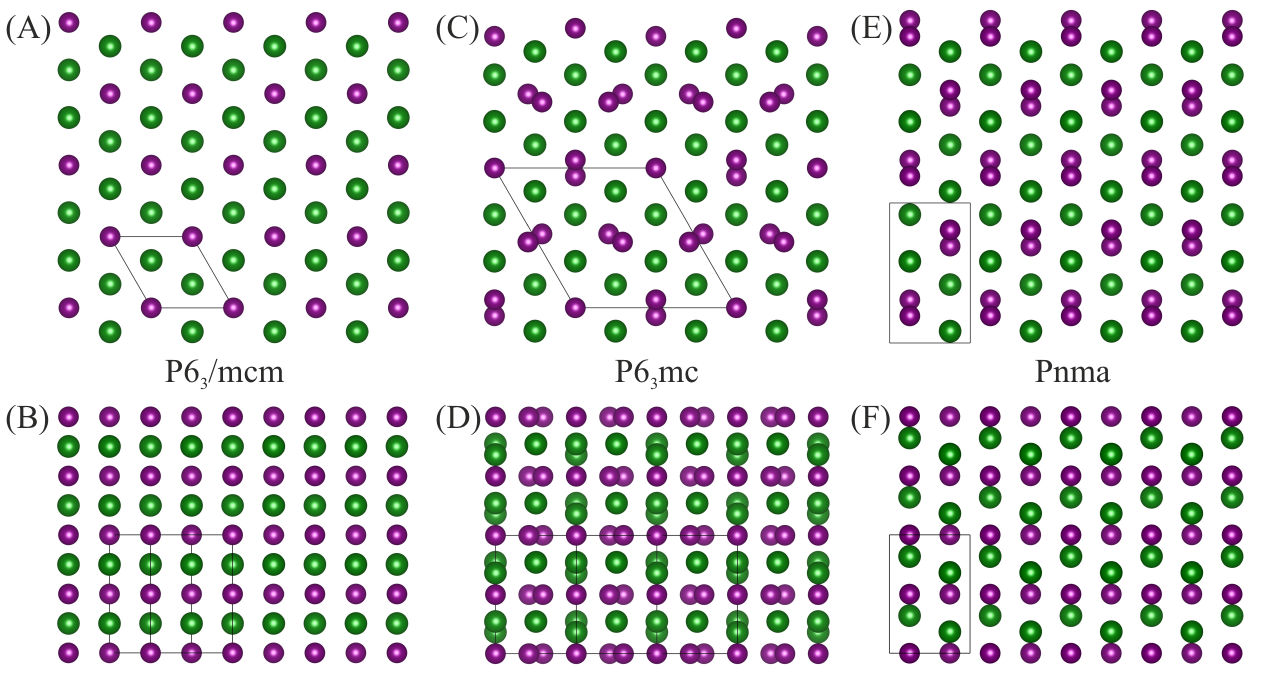}
    \caption{Atomic distortion pattern from the parent non-distorted $P6_3/mcm$ space group (A,B) to $P6_3mc$ (C,D) and $Pnma$ (E,F), including in-plane displacements of Cr atoms (purple) and out-of-plane displacements of Sb atoms (red). The upper row shows the hexagonal $ab$-plane projection, while the lower row shows the out-of-plane projection.}
    \label{SI_superstructure}
\end{figure}

In contrast to the Bragg reflections, the DS is located between the structural Bragg peaks and cannot be indexed using integer Miller indices. Instead, it is concentrated in reciprocal space with propagation vector $\mathbf{k} = (1/2, 0, 0)$, corresponding to a leading instability at the M point of the Brillouin zone. The diffuse pattern is highly symmetric and demonstrates a pronounced dependence on the out-of-plane momentum component. This behavior unambiguously indicates short-range superstructural correlations associated with a locally enlarged unit cell and reduced symmetry. The absence of pronounced DS in the HK0 plane indicates that the atomic displacements possess a predominant out-of-plane component. At the same time, the anisotropy of diffuse spots observed in different HK planes implies the presence of non-zero in-plane atomic displacements.

\begin{figure*}
    \centering
    \includegraphics[width=\linewidth]{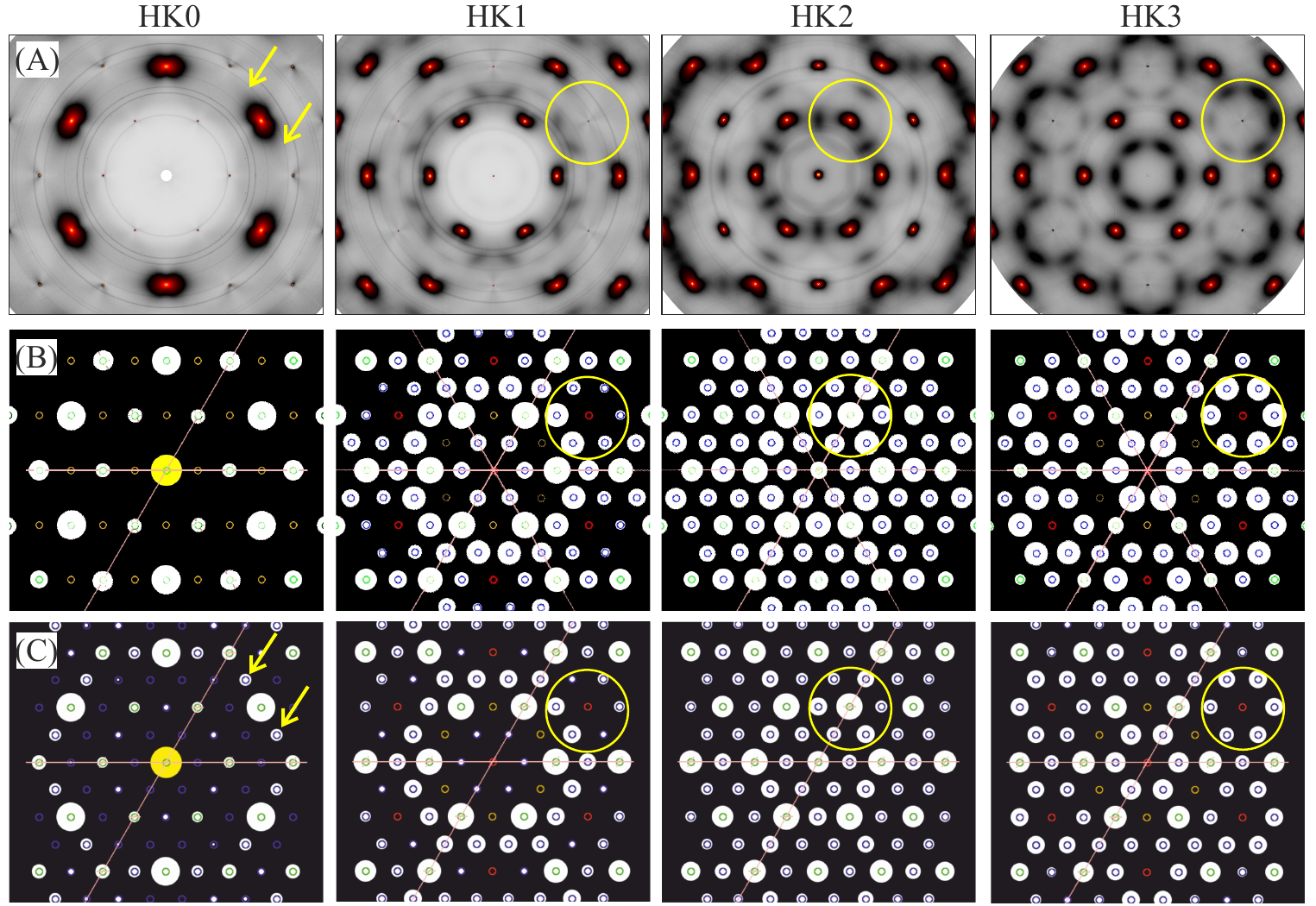}
    \caption{Comparison of the experimental DS pattern (A) measured at 1023~K with patterns generated for distorted structures corresponding to the orthorhombic Pnma (B) and hexagonal P6$_3$mc (C) space groups, depicted in Fig.~\ref{SI_superstructure}.  The simulated maps for the orthorhombic case are symmetrized assuming three equally populated in-plane domains. Yellow circles highlight the same reciprocal-space regions containing characteristic DS features in each map. Yellow arrows indicate additional reflections appearing in the HK0 plane for the hexagonal-type distortion.}
    \label{SI_isodistort}
\end{figure*}

Group--subgroup symmetry analysis shows that several hexagonal and orthorhombic distortions can be induced by an M-point instability of the parent structure. These distortions can be classified according to the relative in-plane and out-of-plane displacement patterns of the Cr and Sb atoms. A comparison of the calculated structure factors with the experimental DS identifies two candidate symmetries that provide the best agreement: hexagonal $P6_3mc$ and orthorhombic $Pnma$ structures, both arising from the irreducible representation M2-. In both cases, the distortion primarily involves in-plane displacements of the Cr atoms accompanied by out-of-plane displacements of the Sb atoms, Fig.~\ref{SI_superstructure}. Representative diffraction patterns generated using ISODISTORT for these two symmetries are shown in Fig.~\ref{SI_isodistort}.

An orthorhombic distortion implies the selection of a preferred in-plane direction. Since no in-plane anisotropy of the DS is observed along the $H00$, $0K0$, and $\overline{H}H0$ directions, this indicates either the preservation of hexagonal symmetry or the presence of three equally populated in-plane orthorhombic domains. Given the short correlation length obtained from the DS, the scenario involving multiple orthorhombic domains is fully plausible. The alternative hexagonal $P6_3mc$ symmetry appears less compatible with the experimental data, as it would allow additional reflections in the $HK0$ plane, which are not resolved, likely due to their low intensity.

As shown in the main text, the high-temperature short-range ordering observed in CrSb is strongly correlated with the suppression of long-range magnetic order, indicating that the associated atomic displacements play a key role in weakening the magnetic exchange interactions mediated by direct Cr--Cr and superexchange Cr--Sb--Cr pathways.

\section{Temperature dependent Hessian phonons and dynamical stability of the non-magnetic phase}

The well known Curie transition towards a non-magnetic phase of CrSb at elevated temperatures, directly implies that such a phase is dynamically stable at those conditions. However, as discussed in the main text, a flat electronic band in the vicinity of the Fermi level induces a Kohn-like anomaly when coupled to the lattice. This phenomena is particular to CrSb, contrary to other similar altermagnetic systems such as MnTe, stable at the harmonic level even without accounting for magnetic order \cite{32gt-3nfn}.

\begin{figure}
    \centering
    \includegraphics[width=0.6\linewidth]{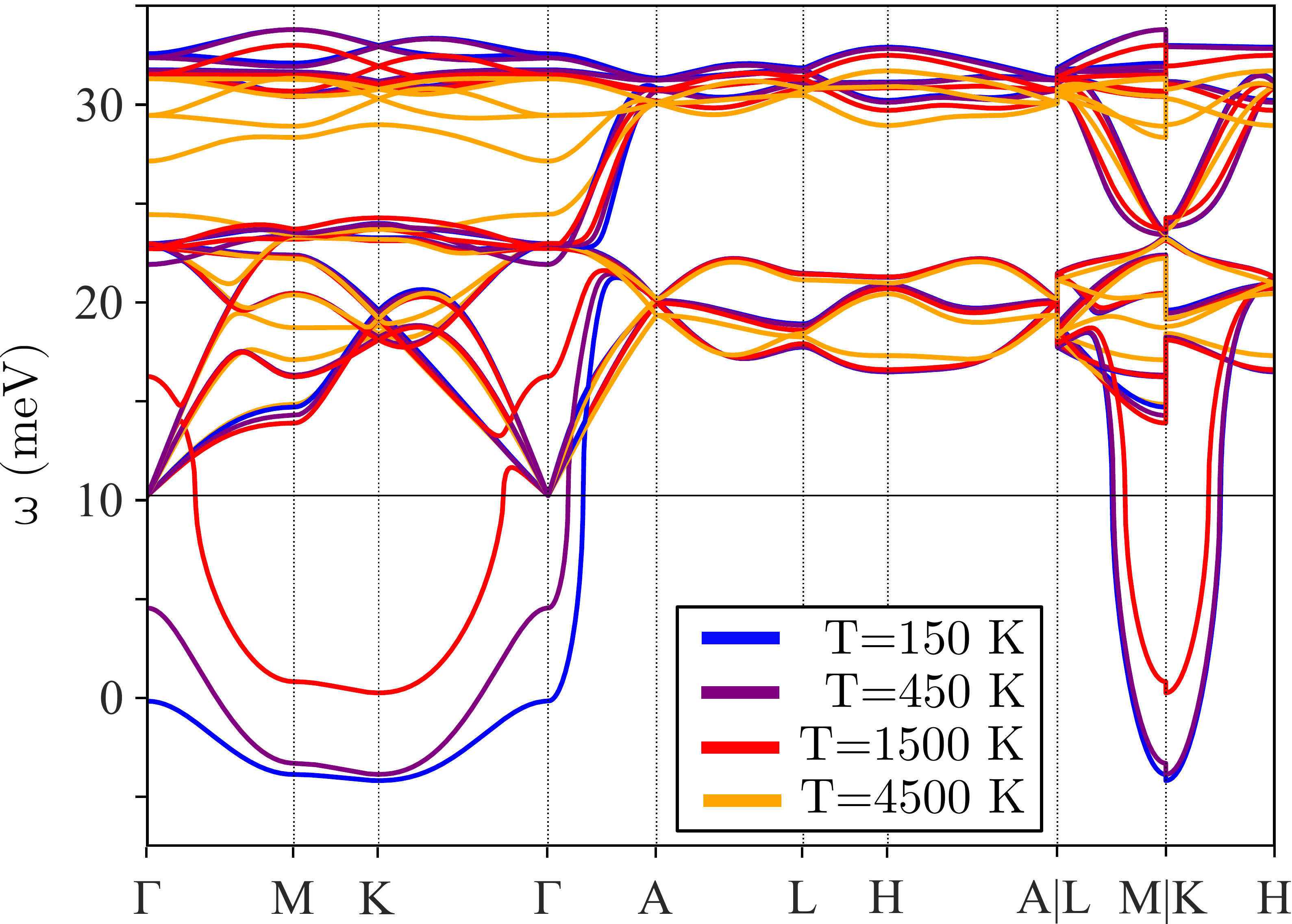}
    \caption{DFPT harmonic phonon spectra as a function of the Fermi-Dirac electronic smearing.}
    \label{fig:S7}
\end{figure}

To account for stabilization, we incorporated electronic smearing using a Fermi-Dirac distribution within the harmonic DFPT framework. However, this resulted in a stabilization temperature of a few thousands of Kelvins, which deviates from the experimental Curie temperature by several orders of magnitude, Fig. \ref{fig:S7}.

Alternatively, we calculated the temperature-dependent Hessian phonon spectra for the non-magnetic phase of CrSb using the Stochastic Self-Consistent Harmonic Approximation (SSCHA) \cite{Errea_2014,PhysRevB.96.014111,Monacelli_2021}, to variationally account for ionic quantum and thermal anharmonic effects, Fig. \ref{fig:S7}. This led to the theoretical prediction of a stable non-magnetic phase above the Curie temperature. The calculations were performed on a $2 \times 2 \times 2$ $\mathbf q$-grid. The K point—the site of the final remaining instability—was not explicitly sampled, but was interpolated for visualization. The excellent quantitative agreement with the IXS measurements presented in the main text reinforces the conclusion that phonon softening in CrSb can be explained solely by ionic quantum and thermal anharmonic effects, albeit coinciding with a purely electronic magnetic transition in temperature, inducing an unprecedentedly large spin-phonon coupling presented in the main text.

\begin{figure}
    \centering
    \includegraphics[width=0.6\linewidth]{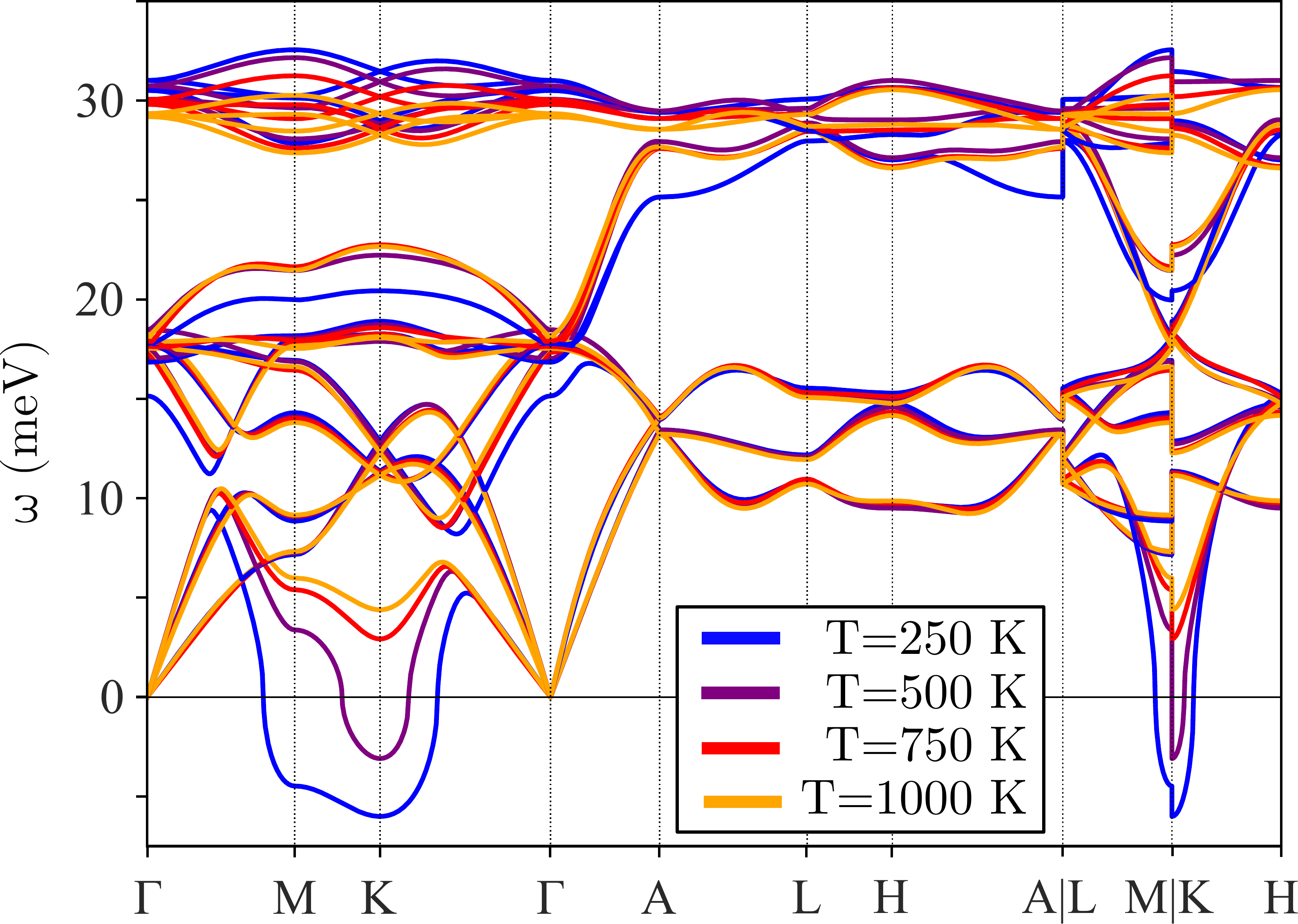}
    \caption{Temperature dependent Hessian phonon-spectra of the non-magnetic phase.}
    \label{fig:S8}
\end{figure}

\section{Projection of the harmonic unstable phonon of the non-magnetic phase into the AFM phonon spectra}

The abrupt stabilization of the lattice accounting for magnetic ordering naturally arises the question whether what occurred with the non-magnetic unstable phonon branch. This is normally done by projecting the polarization vector at a given $\mathbf{q}$-point into the eigenvectors at the magnetic phase, i.e. looking for which vibrational mode $\lambda$, $\boldsymbol{\varepsilon}_{\lambda=0,PM}(\mathbf{q})\cdot\boldsymbol{\varepsilon}^*_{\lambda,AFM}(\mathbf{q})\approx1$. However, such is the impact of spin in the phonon-spectra of the system, that the unstable phonon branch projects over a handful of magnetic vibrational modes depending on the wavevector $\mathbf{q}$, Fig. \ref{fig:S8}.

\begin{figure}
    \centering
    \includegraphics[width=0.6\linewidth]{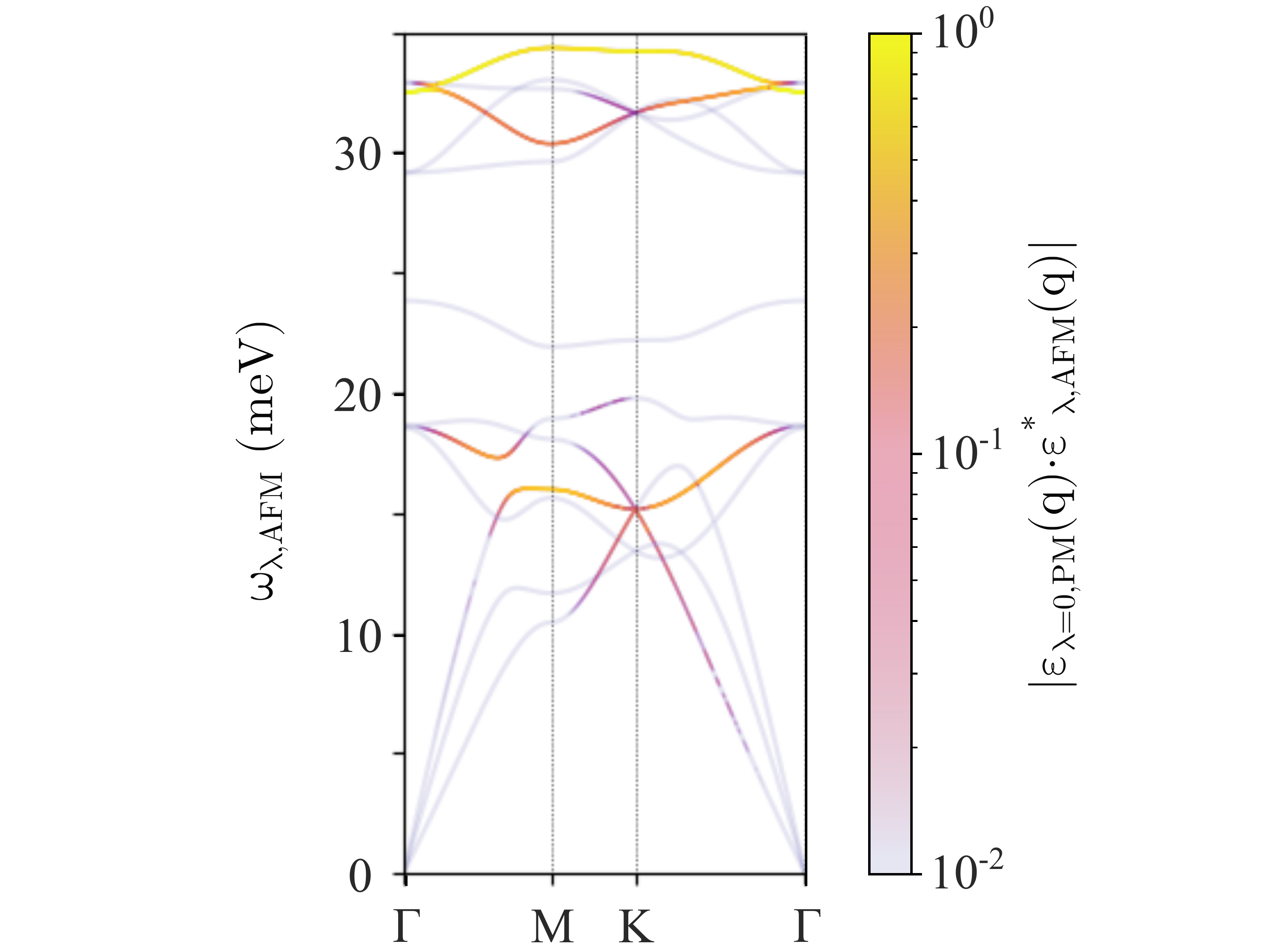}
    \caption{Projection of the DFPT harmonic flat imaginary phonon branch over the altermagnetic harmonic DFPT phonons.}
    \label{fig:S9}
\end{figure}

\section{Calculation of the dynamical structure factor}

To accurately model the IXS measurements, the dynamic structure factor for each vibrational mode must be computed at the specific wave vectors of interest. The structure factor determines how the spectral weight of a mode is distributed in reciprocal space. While this renormalization vanishes at the $\Gamma$ point—where the spectral weight is directly related to the density of states—this is not the case away from the zone center. In these regions, the renormalization of each mode’s spectral weight depends on its polarization vector $\boldsymbol{\varepsilon}_{\lambda}(\mathbf{q})$ and the total momentum transfer $\mathbf{Q} = \mathbf{q} + \mathbf{G}$, where $\mathbf{G}$ is a reciprocal lattice vector.

Following established literature~\cite{cowley_anharmonic_1968, Baron2015}, the one-phonon dynamic structure factor is expressed as:
\begin{equation}
F_\lambda(\mathbf{Q}) = \sum_{a,\alpha} \frac{f_a(\mathbf{Q}) e^{-W_a(\mathbf{Q})}}{\sqrt{2M_a \omega_\lambda(\mathbf{q})}} \left( Q_\alpha \cdot \varepsilon_{\lambda,a}^\alpha(\mathbf{q}) \right) e^{i \mathbf{Q} \cdot \mathbf{R}_a}
\end{equation}
Here, $M_a$ is the mass of atom $a$, $\alpha$ denotes the cartesian index, and $\omega_\lambda(\mathbf{q})$ is the frequency of mode $\lambda$. The term $f_a(\mathbf{Q})$ represents the atomic form factor—the Fourier transform of the atomic electron density—which describes X-ray scattering as a function of momentum transfer $\mathbf{Q}$. The Debye–Waller factor, $e^{-W_a(\mathbf{Q})}$, accounts for the reduction in Bragg intensity due to thermal vibrations, where the exponent is defined as $W_a(\mathbf{Q}) = \frac{1}{2} \langle (\mathbf{Q} \cdot \mathbf{u}_a(\mathbf{q}))^2 \rangle$. Finally, $\mathbf{R}_a$ refers to the equilibrium atomic positions.

To simplify this expression for practical computation, we adopt two primary approximations. First, we assume the atomic form factor is dispersionless and uniform across all species ($f_a(\mathbf{Q}) \approx 1$). Second, we assume that ionic displacements are sufficiently small such that the Debye–Waller factor can be neglected ($e^{-W_a} \approx 1$). Under these assumptions, the dynamic structure factor reduces to:
\begin{equation}
F_\lambda(\mathbf{Q}) = \sum_{a,\alpha} \frac{1}{\sqrt{2M_a \omega_\lambda(\mathbf{q})}} \left( Q_\alpha \cdot \varepsilon^\alpha_{\lambda,a}(\mathbf{q}) \right) e^{i \mathbf{Q} \cdot \mathbf{R}_a}
\end{equation}
This simplified form is the expression employed throughout this manuscript to evaluate the IXS intensities. 

\bibliography{CrSb}